\documentclass[aps,pra,amsmath,amssymb,reprint,floatfix]{revtex4-1}

\usepackage[caption=false]{subfig}
\usepackage{graphicx}
\usepackage{cleveref}
\usepackage{color}

\begin{document}

\title{Localization to delocalization transition in a driven nonlinear cavity array}
\author{Oliver T. \surname{Brown}}
\email{ob25@hw.ac.uk}
\author{Michael J. \surname{Hartmann}}
\affiliation{SUPA, Institute of Photonics and Quantum Physics, Heriot-Watt University, Edinburgh EH14 4AS, United Kingdom}
\date{\today}

\begin{abstract}
We study nonlinear cavity arrays where the particle relaxation rate in each cavity increases with the excitation number. We show that coherent parametric inputs can drive such arrays into states with commensurate filling that form non-equilibrium analogs of Mott insulating states. We explore the boundaries of the Mott insulating phase and the transition to a delocalized phase with spontaneous first order coherence. While sharing many similarities with the Mott insulator to superfluid transition in equilibrium, the phase-diagrams we find also show marked differences. Particularly the off diagonal order does not become long range since the influence of dephasing processes increases with increasing tunneling rates.
\end{abstract}

\maketitle

\section{\label{sec:0} Introduction}
Photons are not usually conserved in light-matter interactions. Consequently, there is no chemical potential for photons, and the rich vein of many-body quantum effects in equilibrium systems is seemingly lost to photonics. Some exceptions, where the concept of an effective chemical potential can be meaningfully applied to photons, include photon emission in semiconductors \cite{Wurfel82}, photons confined in a cavity that couple to excitons and thermalize \cite{KMSL07,EL01,CC13}, and photons interacting with a nonlinear medium that form a Bose-Einstein condensate \cite{Kasprzak:2006fv,KSVW10}. Moreover, in recent years, settings where light-matter interactions can mediate strong photon-photon interactions have gained significant interest as these allow one to generate matter-like phases such as photonic fluids \cite{CC13,VRMea15} or even strongly correlated phases \cite{Hartmann08,Hartmann16,Noh:2017rw}. 

Since photons are bosons, a key question for many-body phenomena in strongly interacting photons or polaritons is whether a phase transition from a Mott insulator to a superfluid state \cite{HP07}, as in  Bose-Einstein condensates \cite{FWGF89,JBCGZ98,GMEHB02} can be observed. Early theory investigations of the phase diagrams of interacting photons or polaritons in arrays of coupled cavities considered equilibrium scenarios by introducing a chemical potential, the physical realization of which remained an open question \cite{Greentree2006,KH09,LOD15,Hartmann16}.
Given the limited lifetime of photons trapped in a cavity, it is however more natural to explore many-body phases in a non-equilibrium setting taking into account input drives and dissipation. Following this route, auxiliary systems together with specific driving mechanisms have recently been considered to generate effective chemical potentials for photons \cite{HAT15,MOHSS17,LBSRFCC17} and resulting phase diagrams have been explored \cite{BSLea17}. 

Here we show that a Mott phase can be generated in a nonlinear cavity array with dissipation by only employing a coherent parametric drive that is directly applied to the cavities and explore the transition from this Mott insulator to a delocalized phase showing first order coherence between lattice sites. 

A key feature of a Mott insulating phase is that the particle number is commensurate with the number of lattice sites, i.e. there is an integer number of particles on each lattice site and the number fluctuations are strongly suppressed. Such a situation cannot be achieved in a nonlinear resonator array with coherent driving at the excitation frequency. In the limit of very strong nonlinearity and vanishing hopping, where the Mott insulating regime is expected, each lattice site may be approximated as a two level system where population inversion corresponding to unit filling cannot be generated via a coherent drive. 

Difficulties in arranging for a commensurate filling at vanishing hopping are not the only challenge for exploring Mott insulating to superfluid transitions in driven dissipative systems. For coherent driving fields, the phase relation between the inputs at different lattice sites is necessarily fixed. Therefore any phase-coherence between light fields in distant cavities that is found can be attributed at least in part to the coherent input drives \cite{Ruiz-Rivas:2014qf} and it is not clear whether such coherence forms spontaneously as in equilibrium \cite{FWGF89,JBCGZ98,GMEHB02}. 

\begin{figure}[ht]
	\includegraphics[width=.7\linewidth]{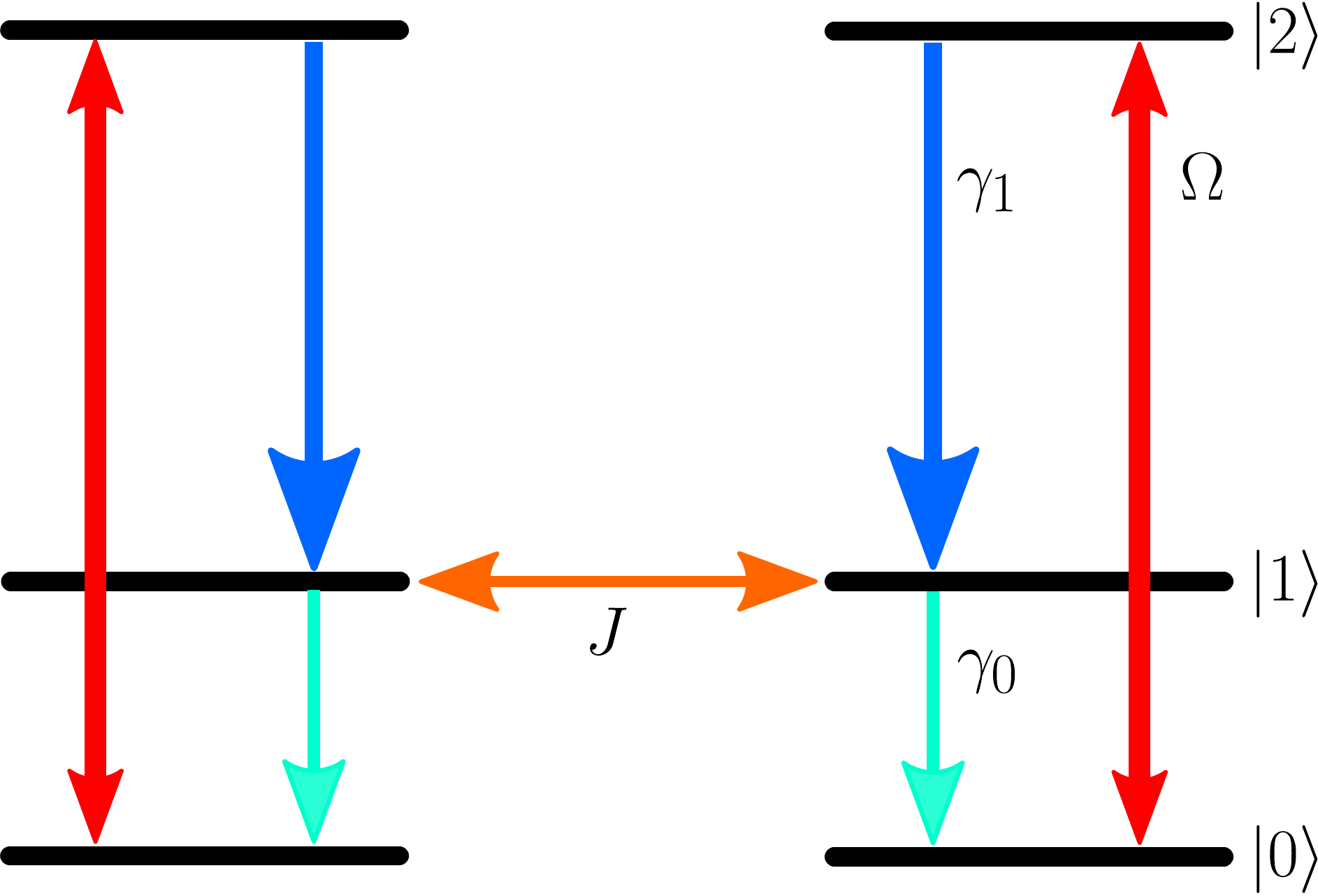}
	\caption{\label{fig:1-1} A diagram of the two-site model, showing states with zero, $|0\rangle$, one, $|1\rangle$ and two, $|2\rangle$, excitations in each cavity as well as the key parameters. The two sites are coupled by a hopping rate, \(J\), there is a coherent parametric drive on each site, \(\Omega\), and there are two dissipative transitions with rates, \(\gamma_{1} \gg \gamma_{0}\).}
\end{figure}

To circumvent the obstacles impeding the formation of Mott insulating phases and the difficulties in studying the spontaneous formation of coherences in coherently driven cavity arrays, we here consider a coherent parametric drive, that resonantly drives the transition from the zero excitation to the two excitation state in each cavity \cite{MOHSS17}, but is off resonance with all other transitions. Together with a cascade of decay processes, where the decay from a two excitation state to a single excitation state, \(\gamma_{1}\), is much faster than the decay of a single excitation state to a zero excitation state, \(\gamma_{0}\), this leads to a stationary state with a very high probability to find a single excitation in each lattice site, see \Cref{fig:1-1} for a sketch of a two-site model. More precisely, the probability to find a single excitation in each lattice site approaches unity in the limit of \(\gamma_{0}/\gamma_{1} \to 0\). 

For this arrangement, we investigate the transition from this Mott phase with commensurate filling to a delocalized phase with incommensurate filling. An important property of our model is that the fast decay at rate \(\gamma_{1}\) erases all coherence between lattice sites that is due to the coherent input. Any first-order correlations between the lattice sites that we find in the stationary states can thus clearly be attributed to the formation of a superfluid component. 

In the following, we first introduce the model we considered, then present our results and finish with conclusions and an outlook.
 
\section{\label{sec:1}The Model}
We consider a system of \(N\) coupled non-linear cavities in a one-dimensional array, governed by a Bose-Hubbard Hamiltonian, with an additional coherent parametric driving term. After moving into a suitable rotating frame, applying the rotating wave approximation and setting \(\hbar = 1\) we are left with the Hamiltonian
\begin{equation}
	\mathcal{H} = \mathcal{H}_{0} + \mathcal{H}_{J} + \mathcal{H}_{\Omega},
	\label{eq:1-1}
\end{equation}
where,
\begin{align}
	\mathcal{H}_{0} &= \sum_{j} \left[ \Delta \hat{a}_{j}^{\dagger}\hat{a}_{j} + \frac{U}{2} \hat{a}_{j}^{\dagger}\hat{a}_{j}^{\dagger}\hat{a}_{j}\hat{a}_{j} \right], \label{eq:1-2} \\
	\mathcal{H}_{J} &= -J \sum_{j} \left[ \hat{a}_{j}\hat{a}_{j+1}^{\dagger} + \hat{a}_{j}^{\dagger}\hat{a}_{j+1} \right], \label{eq:1-3} \\
	\mathcal{H}_{\Omega} &= \sum_{j} \left[ \frac{\Omega}{\sqrt{2}}\hat{a}_{j}^{\dagger}\hat{a}_{j}^{\dagger} + \frac{\Omega^{*}}{\sqrt{2}}\hat{a}_{j}\hat{a}_{j} \right].
	 \label{eq:1-4}
\end{align}
Here \(\Delta = \omega - \omega_{L} / 2\) is the detuning between the drive laser frequency \(\omega_{L}\), and the cavity frequency \(\omega\), \(U\) is the interaction strength, \(J\) is the hopping rate between sites, and \(\Omega\) is the drive strength. We tune the drive laser frequency to be in resonance with the two excitation frequency \(\omega_{L} = 2\omega + U\) which implies \(\Delta = -U/2\).

The dissipative environment we consider is characterized by a cascade of dissipation rates, so the dissipation rate $\gamma_m$ from \(|m+1 \rangle \rightarrow |m \rangle\) is greater than the dissipation rate $\gamma_n$ from \(|n+1 \rangle \rightarrow |n \rangle\) when \(m > n\). We describe this dissipation via a standard Lindblad-form master equation,
\begin{equation}
	\dot{\rho} = -i[\mathcal{H}, \rho] + \sum_{m \geq 0} \mathcal{D}_{m}[\rho],
	\label{eq:1-5}
\end{equation}
where,
\begin{equation}
	\mathcal{D}_{m}[\rho] = \frac{\gamma_{m}}{2} \sum_{j} \left[ 2\kappa_{m,j} \rho \kappa_{m,j}^{\dagger} - \left\{ \kappa_{m,j}^{\dagger}\kappa_{m,j}, \rho \right\} \right]. 
	\label{eq:1-6}
\end{equation}
with jump operators
\begin{equation}
\kappa_{m,j} = |m_j\rangle \langle m+1_j|. 
\label{eq:1-7}
\end{equation}
We note that our model assumes that dissipation is dominated by single particle losses and would reduce to the standard dissipator $\frac{\gamma}{2} \sum_{j} [ 2 a_{j} \rho a_{j}^{\dagger} - \{ a_{j}^{\dagger}a_{j}, \rho \}]$ in the limit where all relaxation rates become equal, $\gamma_m = \gamma$.

We further assume that \( U \gg \Omega \) so that the occupation of levels $|m\rangle$ with $m > 2$ remains negligible and we may truncate our description to the subspace of at most two excitations on each site. It is thus sufficient to ensure that the dissipation rate from \(|2 \rangle \rightarrow |1 \rangle\) exceeds the dissipation rate from \(|1 \rangle \rightarrow |0 \rangle\), i.e. \(\gamma_{1} \gg \gamma_{0}\). Experimentally, such a ratio of dissipation rates can for example be achieved via Purcell enhancement of the relaxation on a specific transition via coupling to a lossy resonator, whose resonance frequency matches this particular transition \cite{Bienfait:2016bs}. 

\section{\label{sec:2}Results}
For the model described by equation (\ref{eq:1-5}) with the Hamiltonian  (\ref{eq:1-1}), we investigated the stationary states. In doing so, we scaled all parameters around the fast dissipation rate, which was fixed at \(\gamma_{1} = 1\). We first consider a small three-site lattice to explore the Mott insulating phase and test the accuracy of our Hilbert space truncation. We then explore the degrading of the Mott insulator and transition to a delocalized phase for larger lattices ($N=11$ and $N=15$) using Matrix Product Operator simulations.

\subsection{\label{sec:2-1}Small anharmonic system}
Before considering the large many-body system we looked at exact calculations for just three sites with periodic boundary conditions, where we extend our description to up to three excitations per site to test the validity of a truncation to two levels. This few site system also allows us to explore the Mott insulating phase where longer range correlations are absent via an exact numerical solution of the model.

The equilibrium phase diagram for the Bose-Hubbard model is typically parameterized by the chemical potential and the hopping between sites. In our case, in contrast, the drive strength and dissipation rates balance out to create an effective chemical potential. For this reason we consider the drive strength and the hopping rate to be an appropriate parameterization to explore the phase transition we are interested in.

\begin{figure*}[ht]
	\subfloat[\label{fig:2-1a}]{\includegraphics[width=0.47\linewidth]{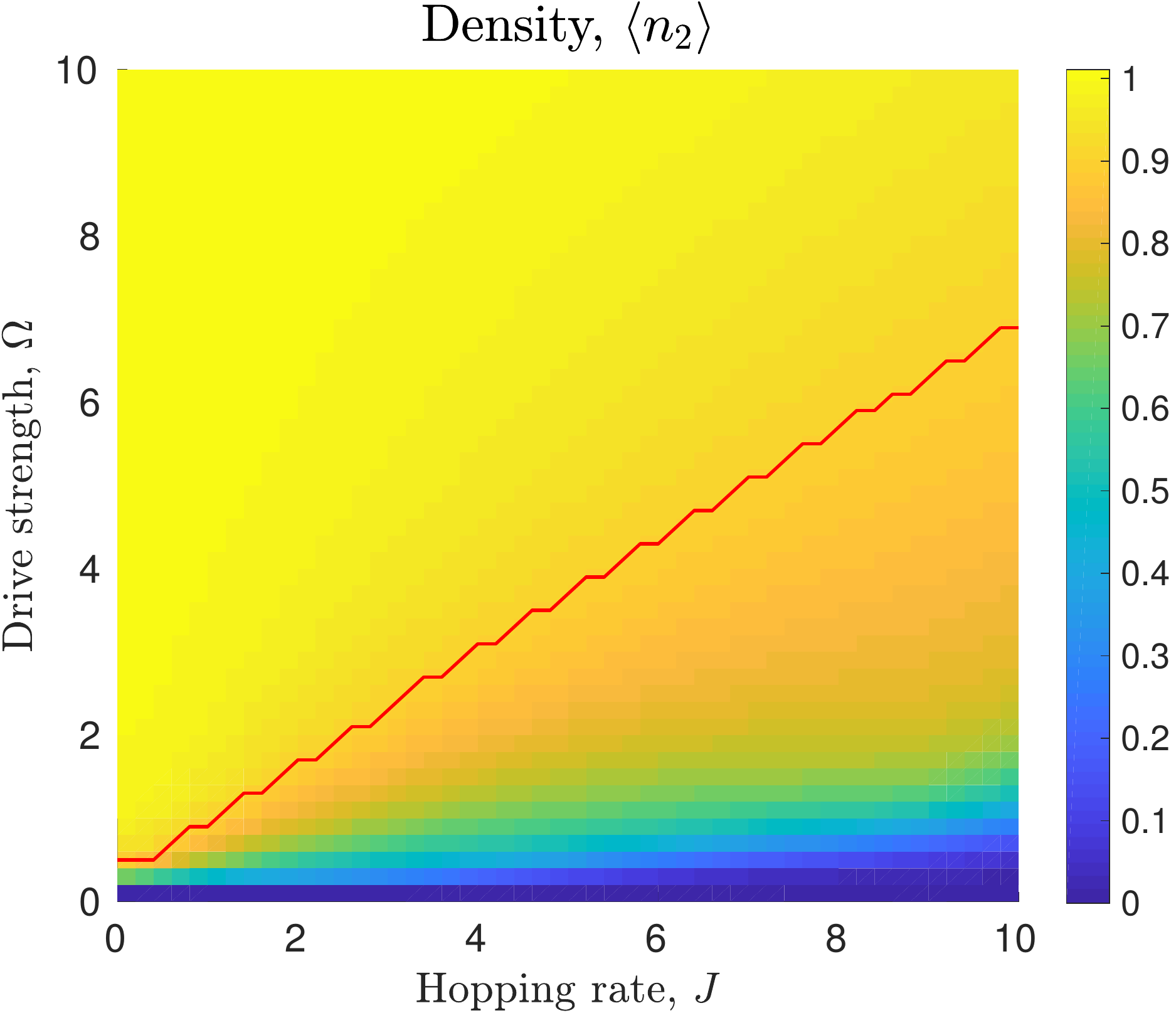}} \hfill
	\subfloat[\label{fig:2-1b}]{\includegraphics[width=0.49\linewidth]{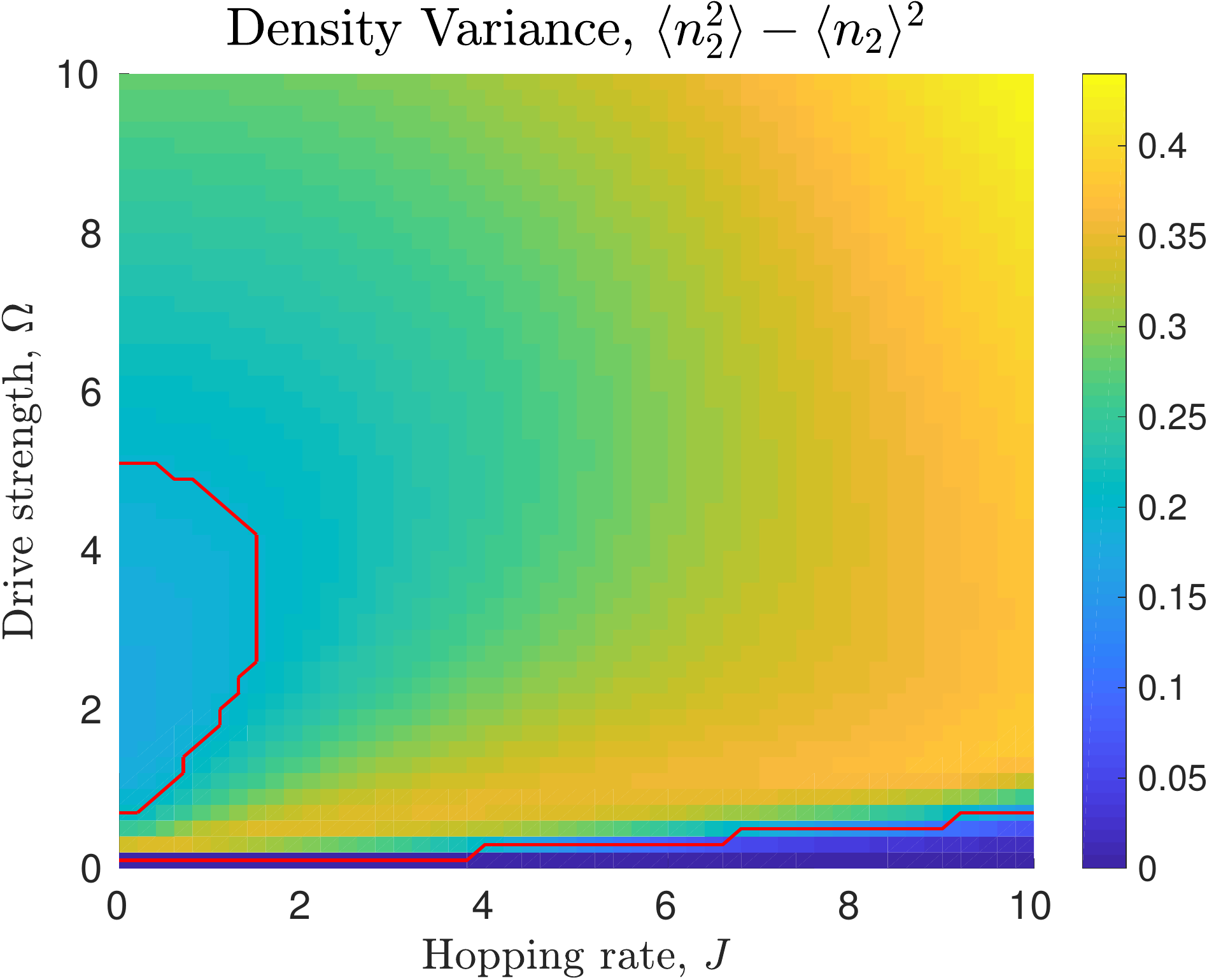}}
	\caption{\label{fig:2-1} (a) The density of the middle site in a three site system, plotted against drive strength \(\Omega\), and coupling strength \(J\). The area above the red line has a density of \(1 \pm 0.1\). (b) The variance, \(\langle n_{2}^{2} \rangle - \langle n_{2} \rangle^{2}\) over the same parameter range. For this calculation, \(\gamma_{2} = 10\), \(\gamma_{1} = 1\),\(\gamma_{0} = 0.1\), \(U=100\), and so \(\Delta = -50\). The area bounded by the red line has a variance of \(\leq 0.2\).}
\end{figure*}

\Cref{fig:2-1} shows the number density \(\langle n_{2} \rangle\) and its variance \(\langle n_{2}^{2} \rangle - \langle n_{2} \rangle^{2} \) for one site, the site with index 2, in this translation invariant system. Both quantities are plotted against the drive strength, \(\Omega\), and the coupling strength, \(J\).
As can be seen from \Cref{fig:2-1}, there is a region, in the shape of a well known Mott lobe, at low hopping rate for which the density is unity with good accuracy, and the variance \(\langle n_{2}^{2} \rangle - \langle n_{2} \rangle^{2} \ll 1\). This shows that there is a stationary state phase for our model with very similar properties as a Mott insulator in equilibrium systems.  

To explore the boundaries of this phase and its transition to a delocalized phase, with first order correlations between lattice sites, we also considered larger lattices with $N=15$, which we discuss next.

\subsection{\label{sec:2-2}Large anharmonic system}
\begin{figure*}[ht]
	\subfloat[\label{fig:2-2-1a}]{\includegraphics[width=0.47\linewidth]{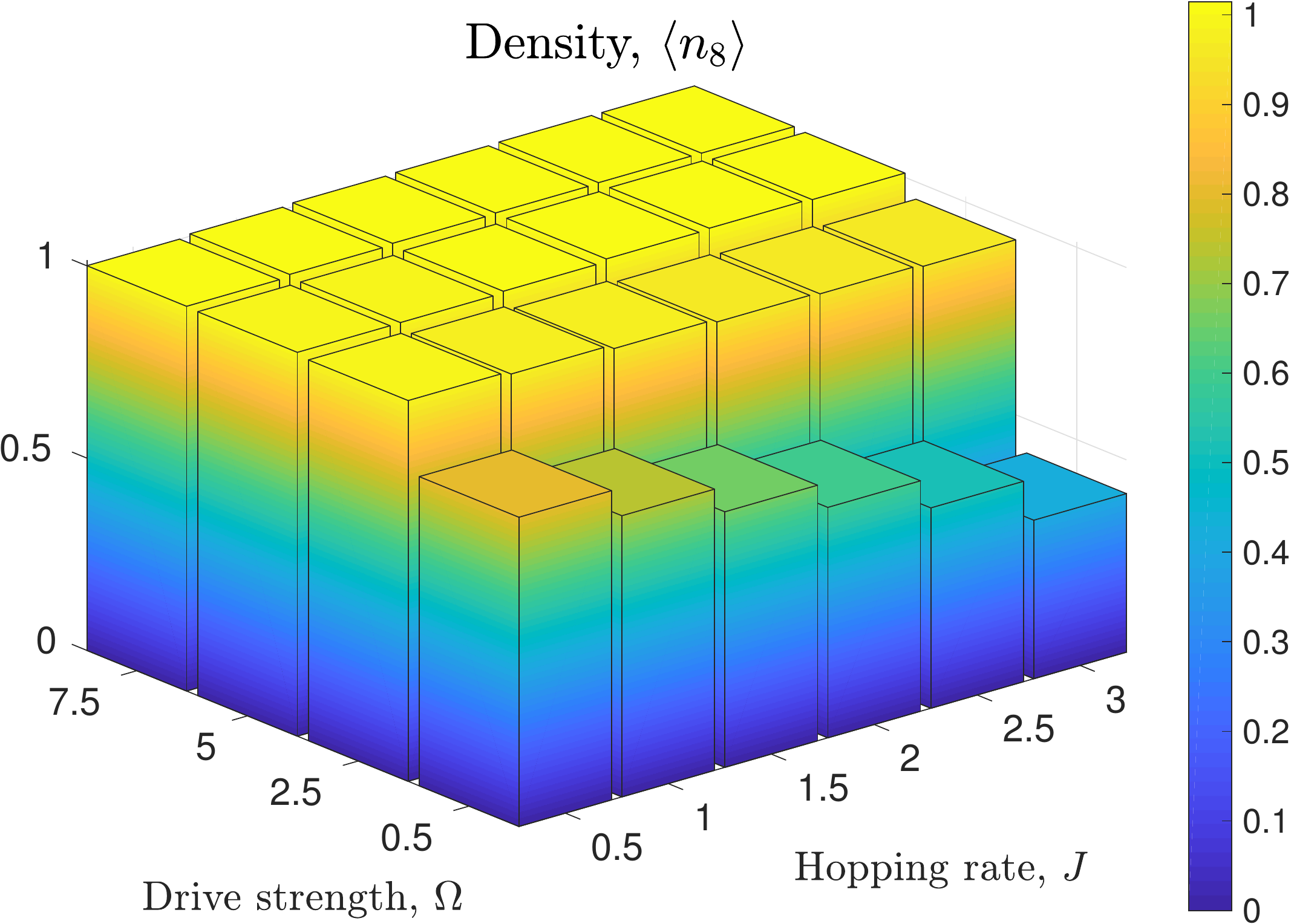}} \hfill
	\subfloat[\label{fig:2-2-1b}]{\includegraphics[width=0.49\linewidth]{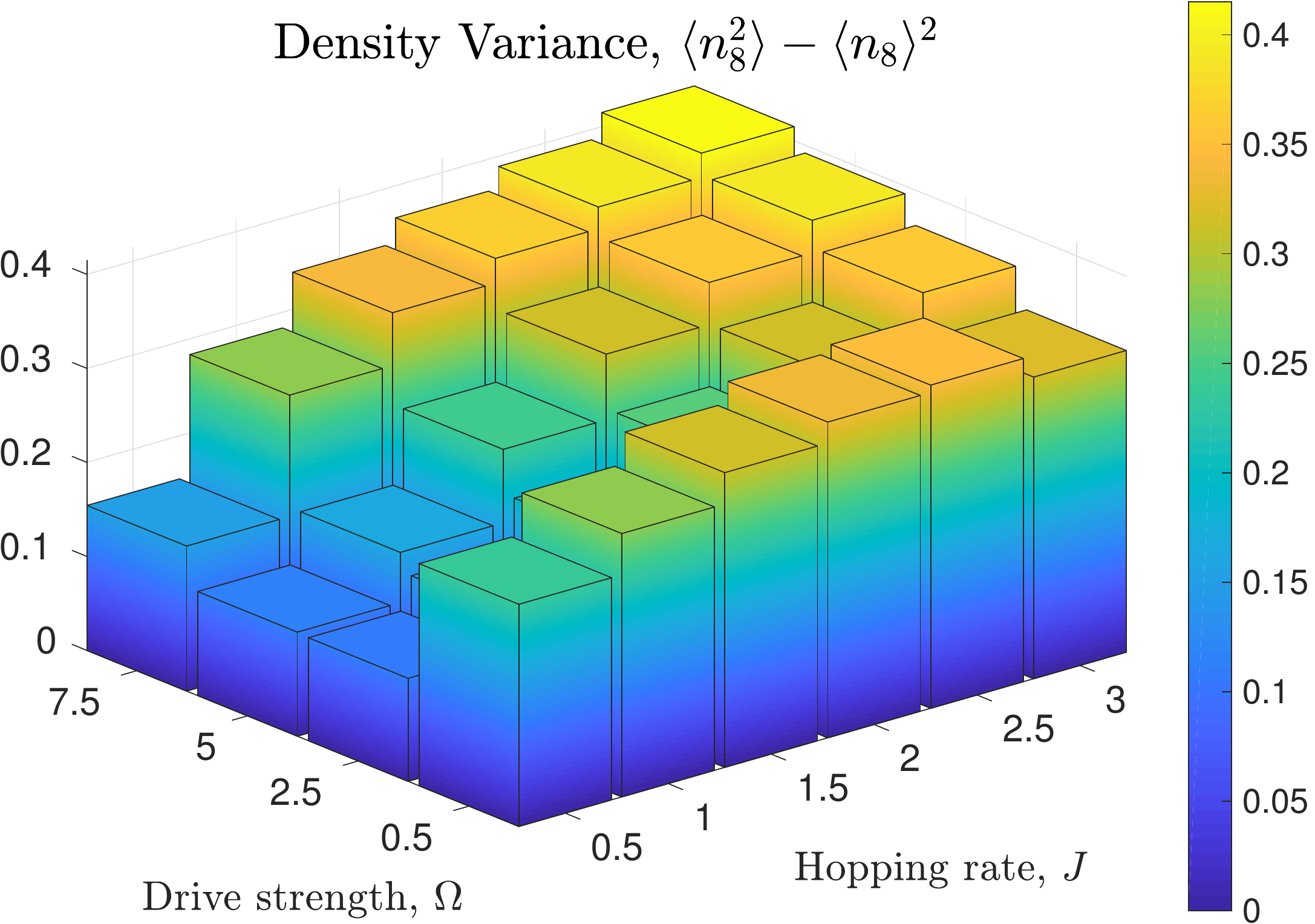}}
	\caption{\label{fig:2-2-1} (a) The density of the middle site in an eleven site system, plotted against drive strength \(\Omega\), and coupling strength \(J\). (b) The variance in the density over the same parameter range. For this calculation \(\gamma_{1} = 1\), \(\gamma_{0} = 0.1\), \(U = 20\), and so \(\Delta = -10\).}
\end{figure*}

In order to explore the build up of correlations and analyze their length, we moved to considering a larger 15-site system, with up to two excitations per site. This was achieved using an implementation of the Time Evolving Block Decimation (TEBD) method \cite{Vidal04}. Unlike the small system calculation, we here consider a system with open boundary conditions.  \Cref{fig:2-2-1} shows the density of the middle site and its variance plotted against drive strength and coupling strength. 
It can be seen that the density remains close to one for sufficient drive strength $\Omega$, and again at low coupling strength $J$ we see a variance of much less than one, indicative of the Mott insulator state. 

For the parameter region where the local excitation number fluctuations start to increase, it is an interesting question whether a transition to a superfluid state or BEC occurs. A signature of such a transition would be an increase in long range first order coherence as found in an equilibrium BEC. We therefore investigated these first order correlations as quantified by the $g^{(1)}$-function, 
\begin{equation}
g^{(1)}(i,j) = \frac{\langle \hat{a}_{i}^{\dagger}\hat{a}_{j} \rangle}{\sqrt{\langle \hat{n}_{i}\rangle \langle \hat{n}_{j} \rangle}}.
\label{eq:2-2-1}
\end{equation}
\begin{figure*}[ht]
	\subfloat[\label{fig:2-2-2a}]{\includegraphics[width=0.49\linewidth]{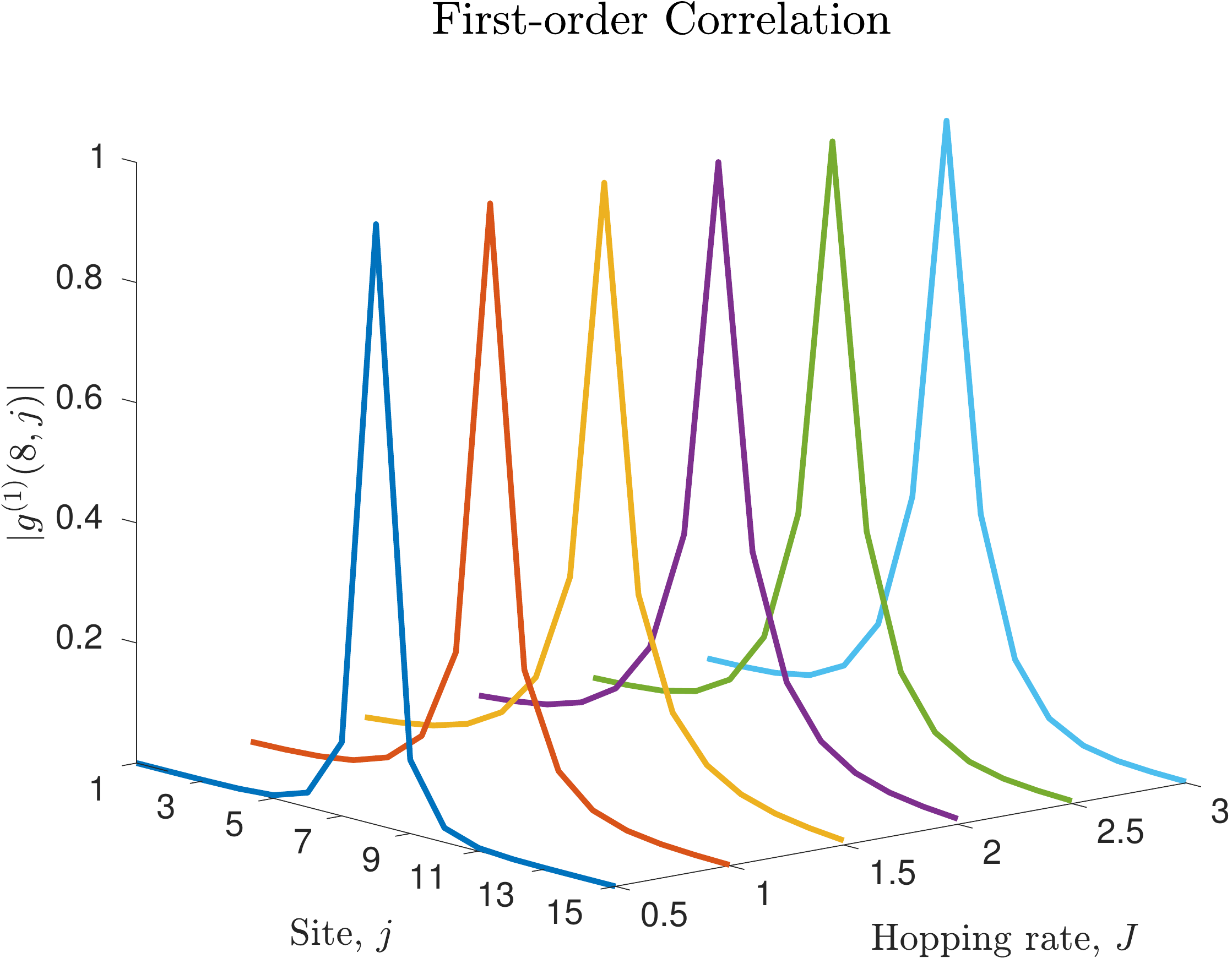}} \hfill
	\subfloat[\label{fig:2-2-2b}]{\includegraphics[width=0.48\linewidth]{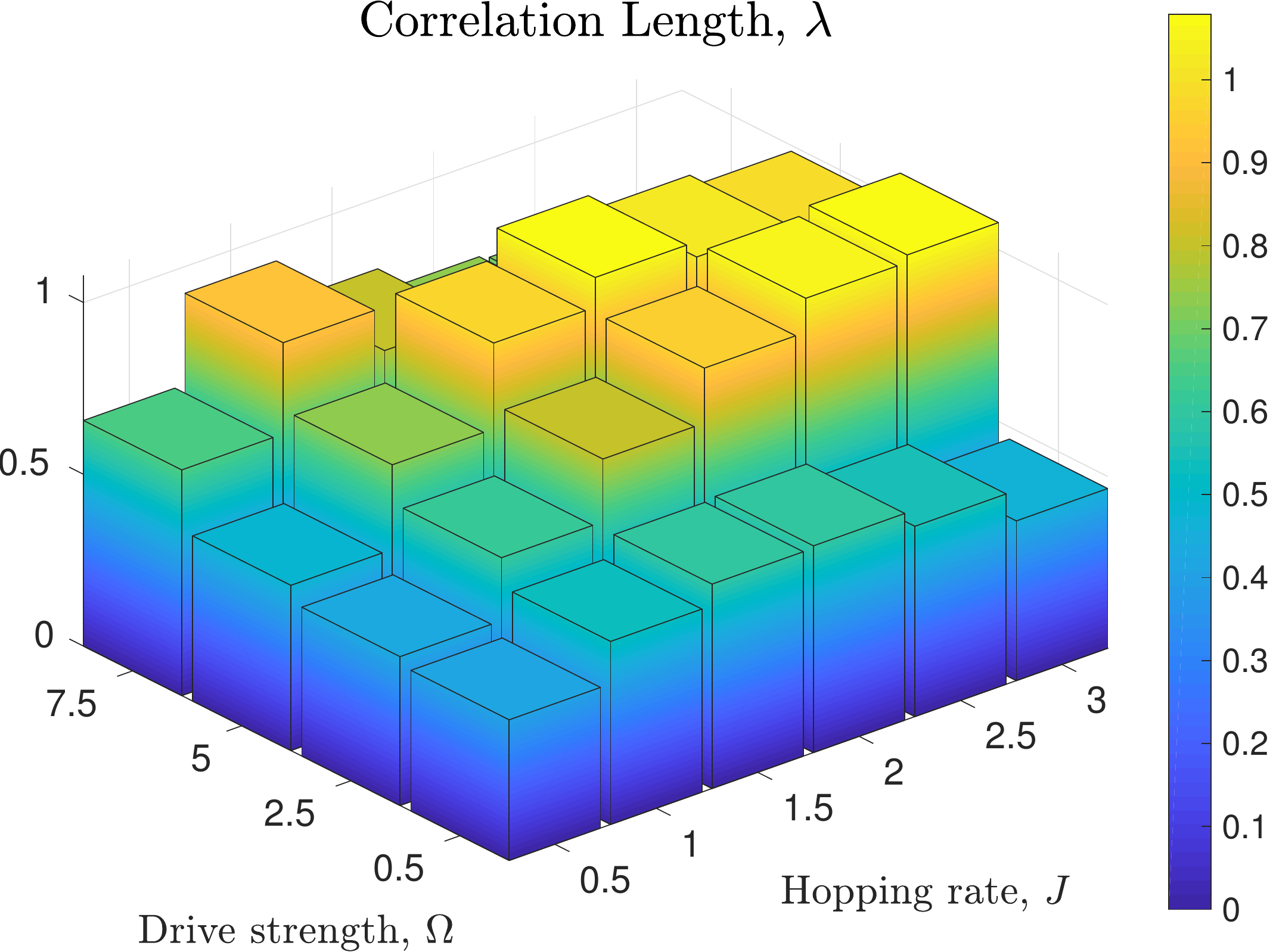}}
	\caption{\label{fig:2-2-2} The first order correlation \(g^{(1)}(i,j)\) and the correlation length $\lambda$. For this calculation, \(\gamma_{1} = 1\), and \(\gamma_{0} = 0.1\). The interaction strength, \(U = 20\), and so \(\Delta = -10\). The first order correlation is plotted for a range of coupling strengths at a fixed drive strength, \(\Omega = 5\). The correlation length was determined by an \(\exp(- |j-j_{0}|/\lambda)\) fit to the \(g^{(1)}\) data.}
\end{figure*}
It can be seen in \cref{fig:2-2-2a} that first order correlations build up in the system as the coupling strength increases. The range of these correlations however does not increase monotonically with $J$, but drops after reaching a peak. To obtain a correlation length we fitted an exponential \(\exp(-|j-j_{0}|/\lambda)\) to the \(g^{(1)}\) data, where $\lambda$ is the correlation length. The result is shown in \cref{fig:2-2-2b}. We attribute the non-monotonic behavior of the correlation length to the competition between the tunneling processes, which enhance long range coherence, and the increase of dephasing processes with increasing local density fluctuations, which reduce long range coherence. Specificly the local density fluctuations $\langle n_{j}^{2} \rangle - \langle n_{j} \rangle^{2}$ increase with the tunneling rate $J$ and cause more occupation of the double excitation levels, which in turn leads to a stronger contribution of the fast relaxation mechanism and thus enhanced dephasing.

\begin{figure}[ht]
	\includegraphics[width=\linewidth]{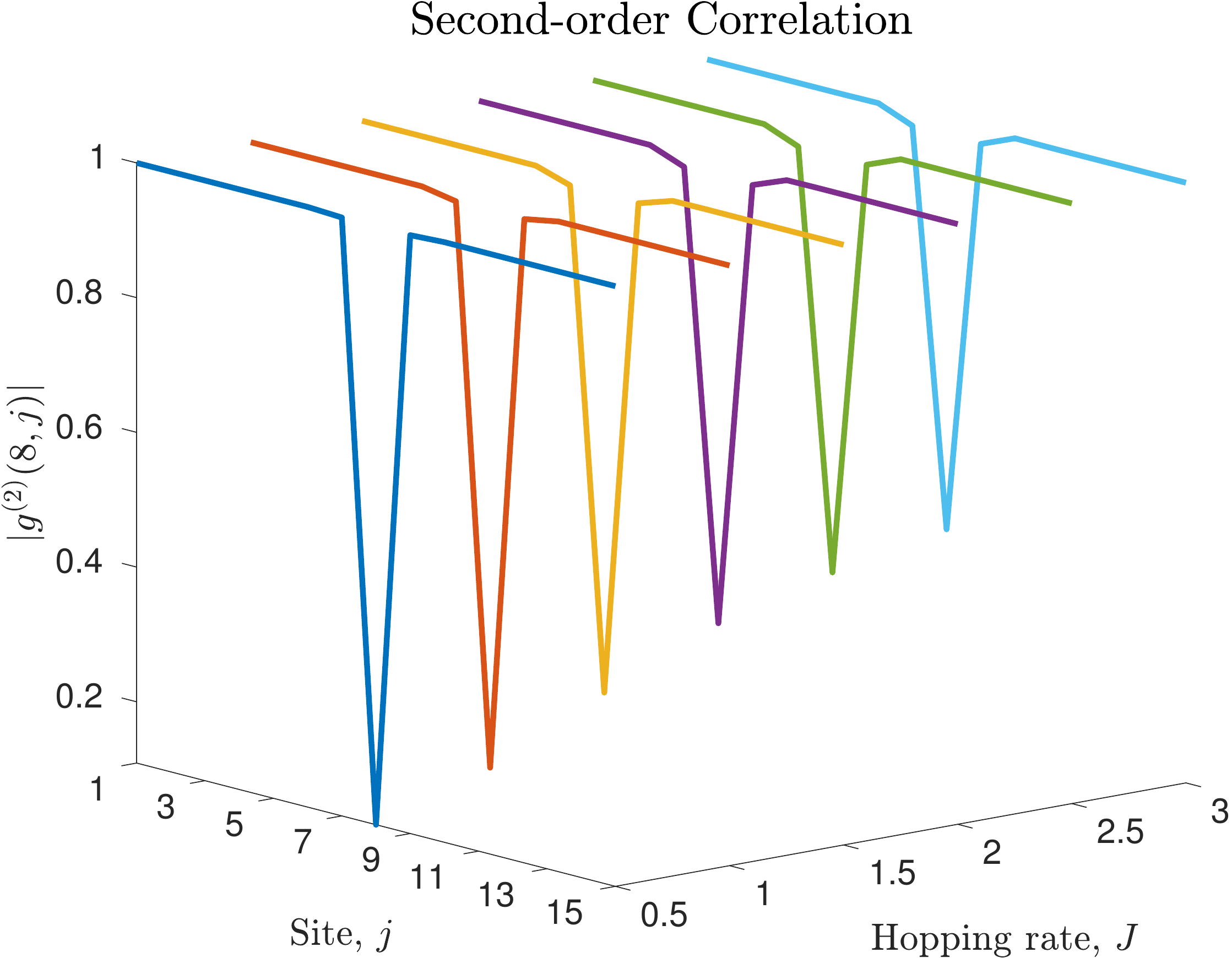}
	\caption{\label{fig:2-2-3} Second order correlation between sites, \(g^{(2)}(i,j)\) plotted against coupling strength. For this calculation, drive strength \(\Omega = 5\), \(\gamma_{1} = 1\), and \(\gamma_{0} = 0.1\). The interaction strength, \(U=20\), and so \(\Delta = -10\).}
\end{figure}

A further characteristic of a Mott insulator is incompressibility.
To explore whether the system becomes more compressible with increasing hopping rate, we thus calculated two-photon coincidences as quantified by the $g^{(2)}$-function, 
\begin{equation}
g^{(2)}(i,j) = \frac{\langle \hat{a}_{i}^{\dagger}\hat{a}_{j}^{\dagger}\hat{a}_{j} \hat{a}_{i}\rangle}{\langle \hat{n}_{i}\rangle \langle \hat{n}_{j} \rangle}.
\label{eq:2-2-2}
\end{equation}
\Cref{fig:2-2-3} shows that indeed the on-site density correlations as measured by $g^{(2)}(j,j)$ increase monotonically across this region.

\subsection{\label{sec:2-3}Large harmonic system}

\begin{figure*}[ht]
	\subfloat[\label{fig:2-3-2a}]{\includegraphics[width=0.49\linewidth]{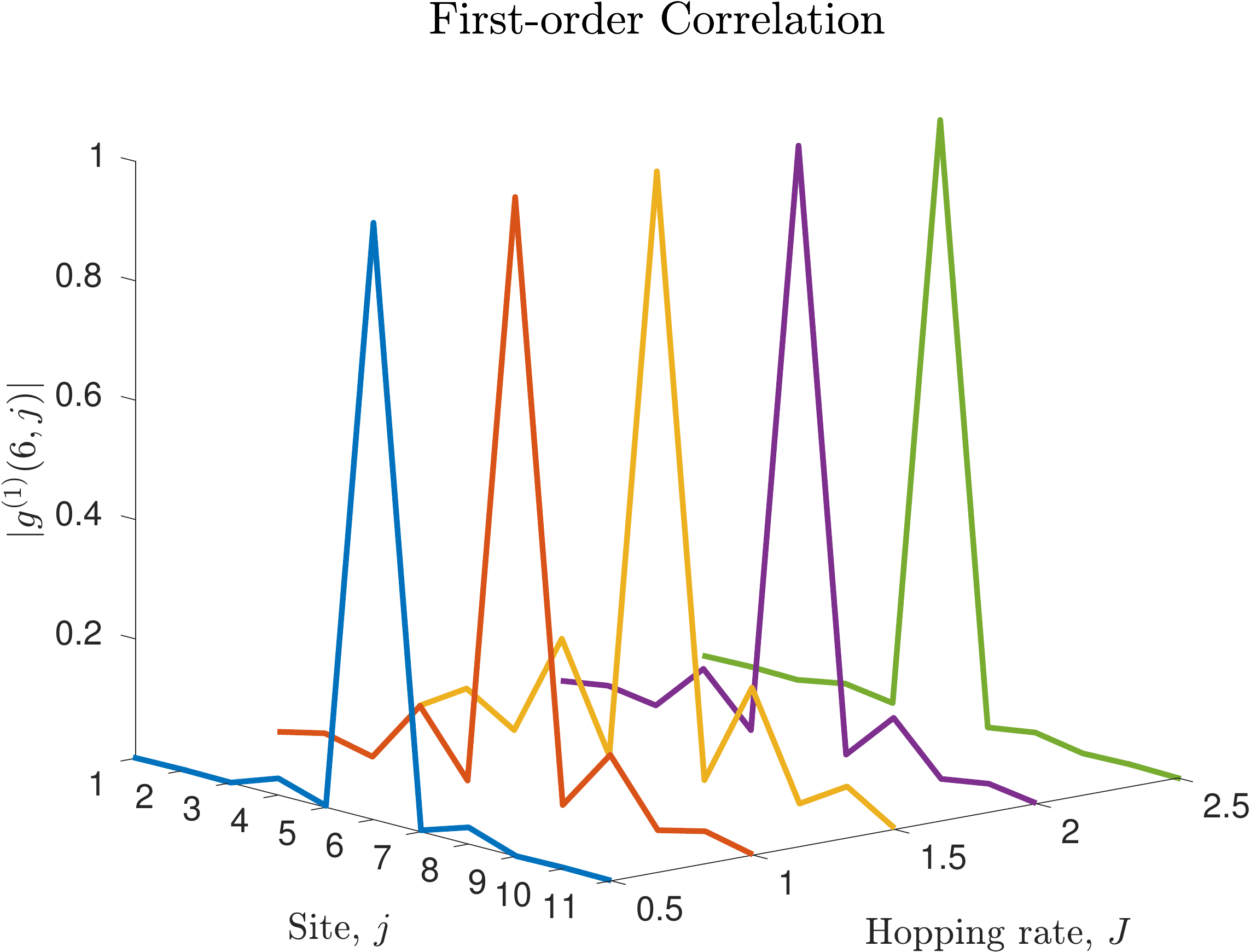}} \hfill
	\subfloat[\label{fig:2-3-2b}]{\includegraphics[width=0.48\linewidth]{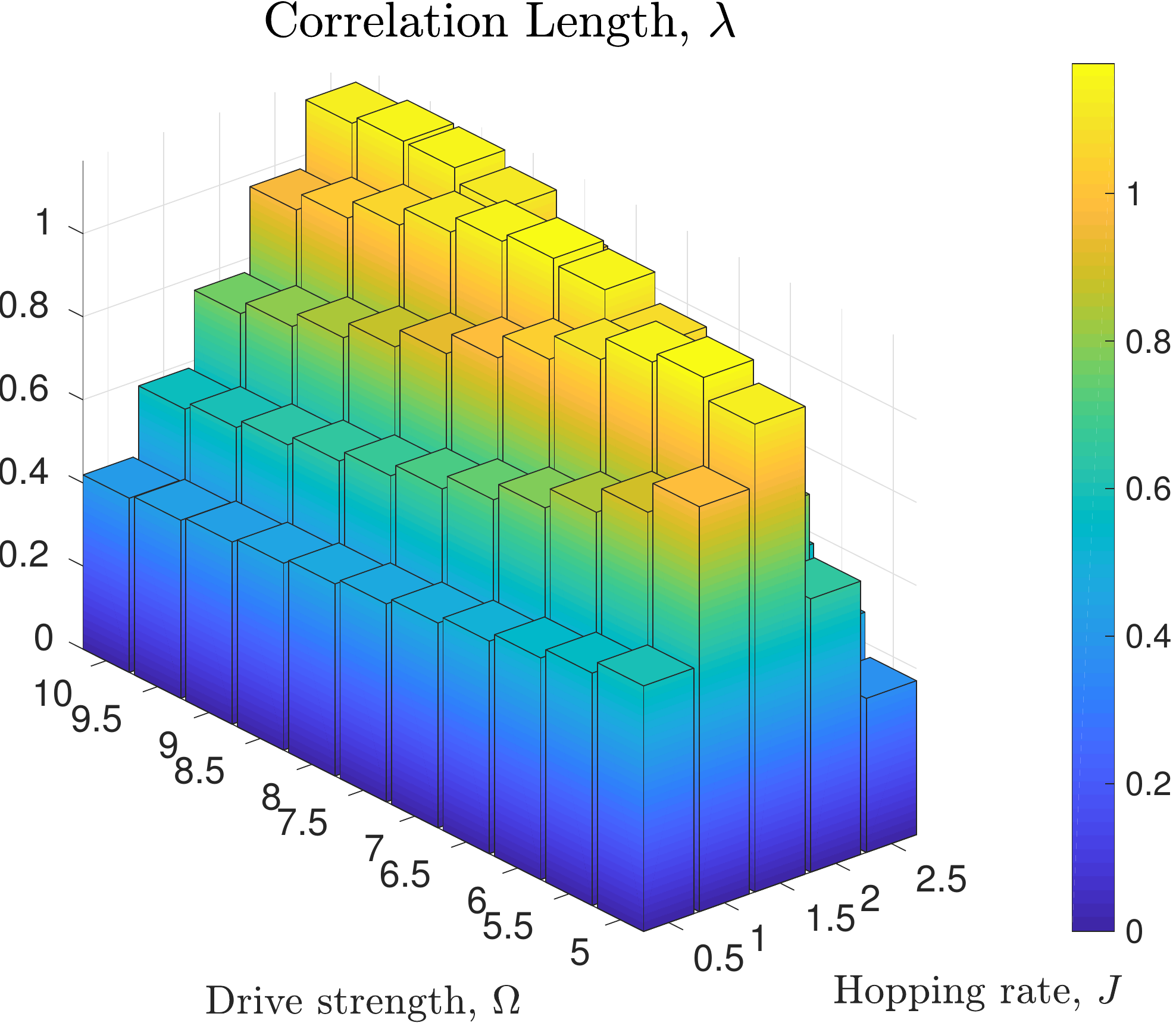}}
	\caption{\label{fig:2-3-2} The first order correlation \(g^{(1)}(i,j)\) and the correlation length $\lambda$. For this calculation, \(\gamma_{1} = 1\), and \(\gamma_{0} = 0.1\). The first order correlation is plotted for a range of coupling strengths at a fixed drive strength, \(\Omega = 5.5\). The correlation length was determined by an \(\exp(- |j-j_{0}|/\lambda)\) fit to the \(g^{(1)}\) data.}
\end{figure*}

\begin{figure}[ht]
	\includegraphics[width=\linewidth]{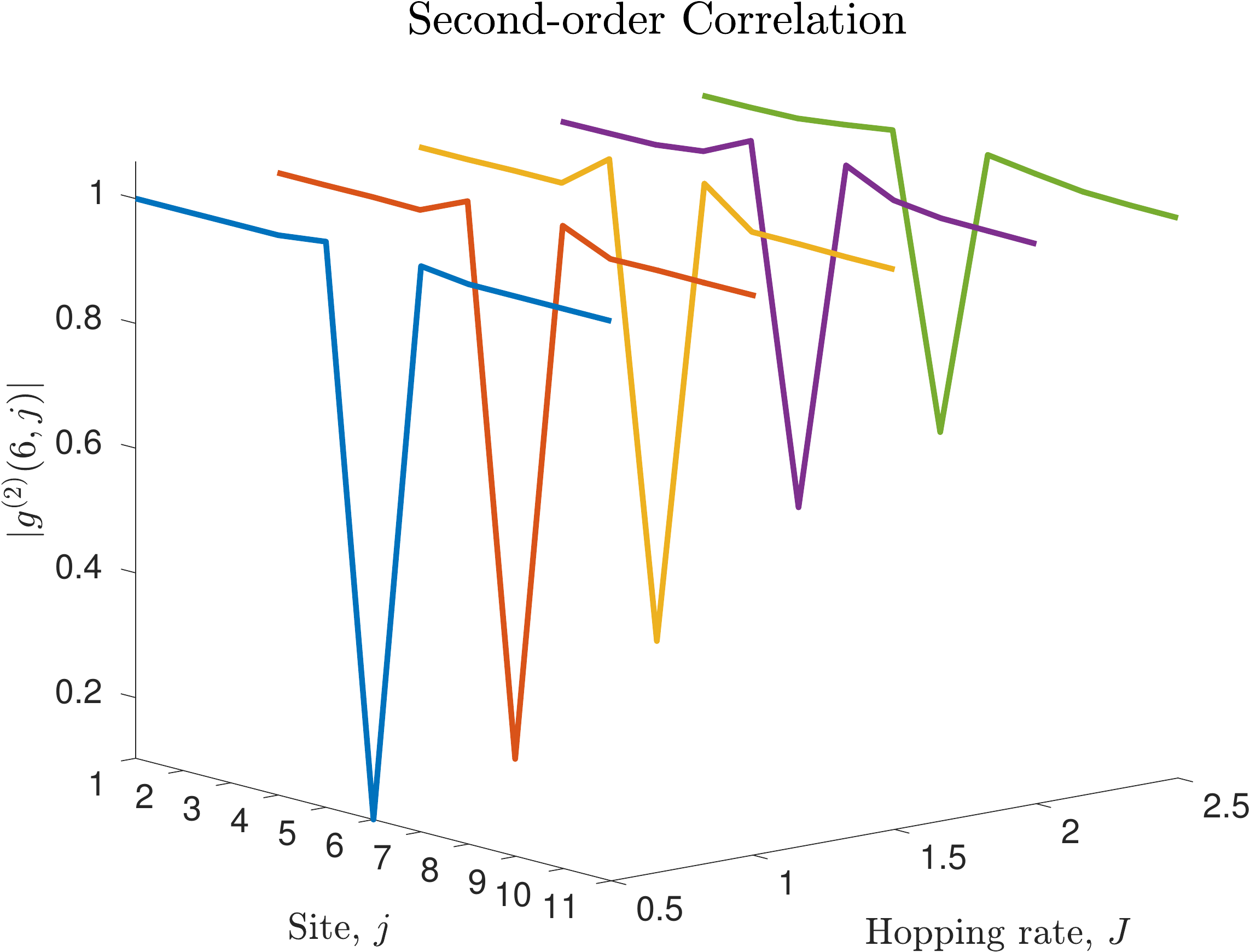}
	\caption{\label{fig:2-3-3} Second order correlation between sites, \(g^{(2)}(i,j)\), plotted against coupling strength. For this calculation, drive strength \(\Omega = 5.5\), \(\gamma_{1} = 1\), and \(\gamma_{0} = 0.1\).}
\end{figure}

To show that our findings are not limited to the parameters considered so far and to further explore this class of systems, we considered a lattice with $\Delta = 0$ and hence $U = 0$. We again used a truncation to the subspace of at most two excitations and calculated the steady state using a TEBD code for an 11-site system with open boundary conditions. 
\begin{figure*}[ht]
	\subfloat[\label{fig:2-3-1a}]{\includegraphics[width=0.47\linewidth]{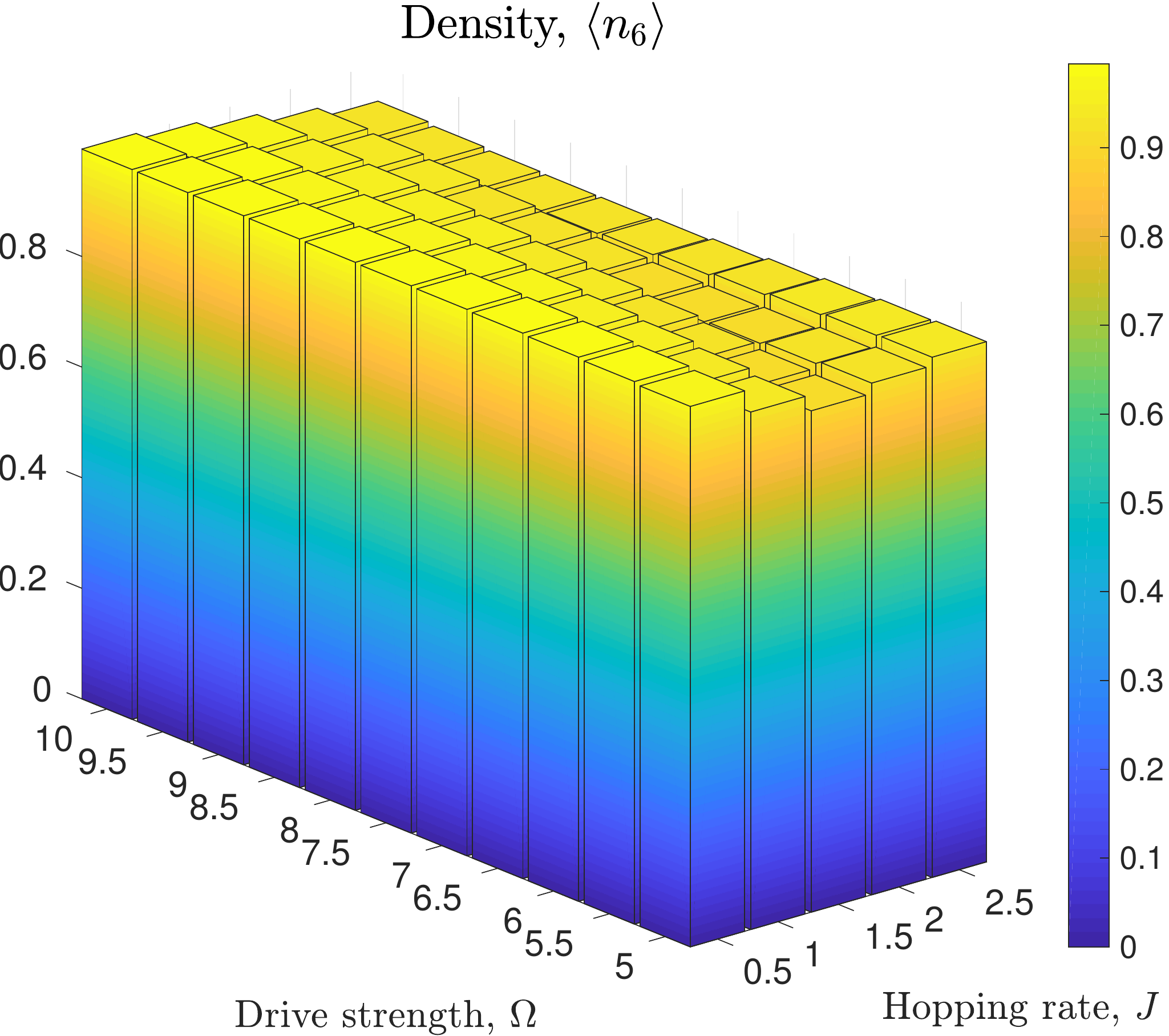}} \hfill
	\subfloat[\label{fig:2-3-1b}]{\includegraphics[width=0.49\linewidth]{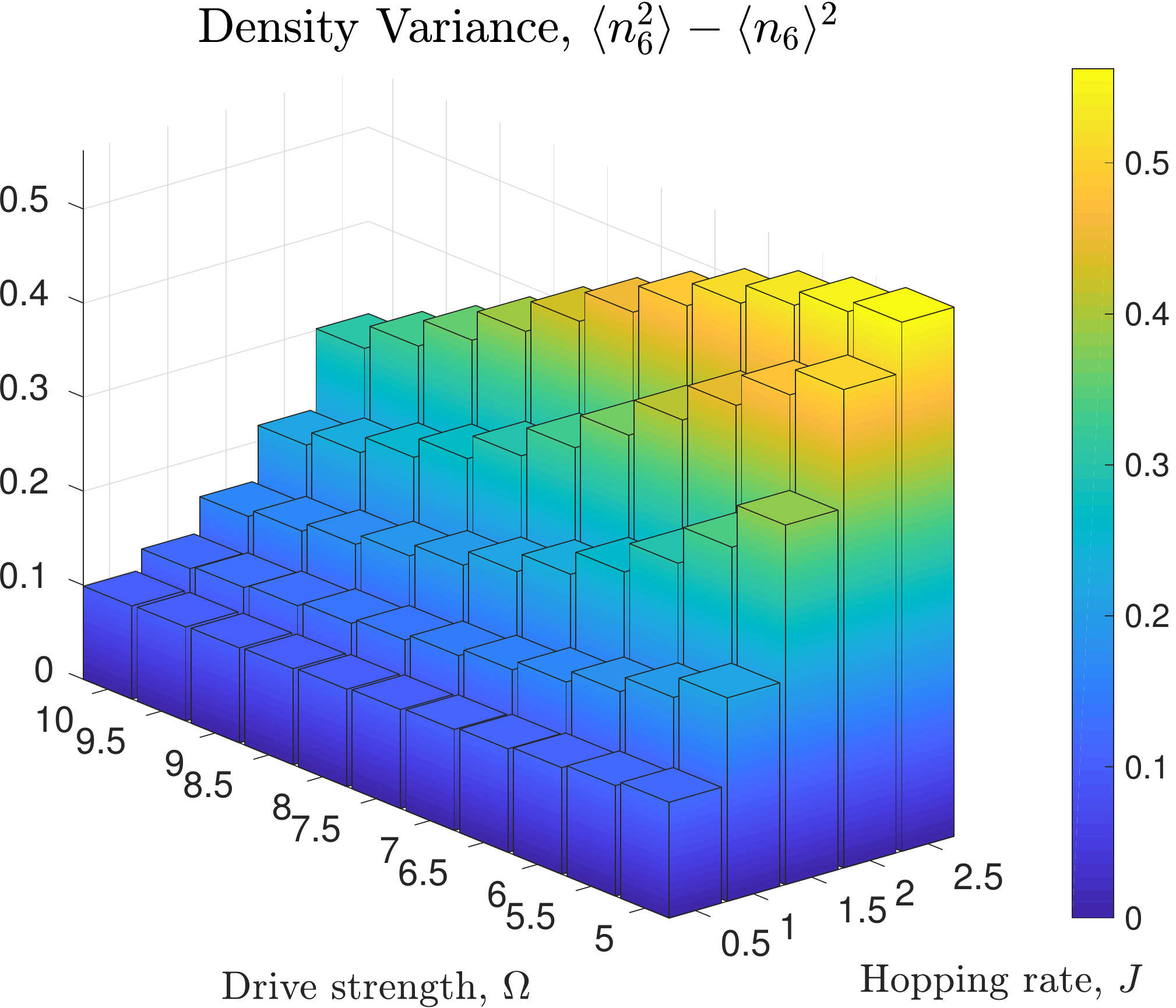}}
	\caption{\label{fig:2-3-1} Plots of the density on the middle site of the system, and it's variance, plotted against drive strength \(\Omega\), and coupling strength \(J\). For this calculation, \(\gamma_{1} = 1\), and \(\gamma_{0} = 0.1\).}
\end{figure*}
\Cref{fig:2-3-1} shows the middle-site density, and density variance plotted against drive and coupling strength. Here we also find a region where the excitation density is approximately commensurate and its variance is much less than one, indicative of the Mott insulator state. This region is however shifted in the parameter space to a higher drive strength. In \cref{fig:2-3-2,fig:2-3-3} we again show the correlation length against drive strength and hopping rate, and the first and second order correlations for a fixed drive strength.

The principle difference between the correlations in the harmonic and anharmonic case is the presence of troughs in the first-order correlation on alternating sites in the harmonic case. This can be understood from the momentum basis representation of the master equation. In this representation, the system is modeled by the Hamiltonian,
\begin{align}
	\mathcal{H} &= \sum_{k=0}^{N-1} \left[ \frac{\Omega}{\sqrt{2}}\hat{b}_{k}^{\dagger}\hat{b}_{N-k}^{\dagger} + \frac{\Omega^{*}}{\sqrt{2}}\hat{b}_{k}\hat{b}_{N-k} \right] \notag \\ 
	&+ \sum_{k=0}^{N-1} \left[\Delta - 2J\cos\left(\frac{2\pi k}{N}\right)\right]\hat{b}_{k}^{\dagger}\hat{b}_{k} \notag \\
	&+ \sum_{(j,k,l,m)} \left[ \frac{U}{N} \hat{b}_{j}^{\dagger}\hat{b}_{k}^{\dagger}\hat{b}_{l}\hat{b}_{m} \right],
	\label{eq:2-3-1}
\end{align}
where $\hat{a}_{n} = \frac{1}{\sqrt{N}} \sum_{k=0}^{N-1} \exp(i\frac{2\pi n}{N}k) \hat{b}_{k}$ and we have assumed periodic boundary conditions. The notation \((j,k,l,m)\) indicates that the indices range from \(0\) to \(N-1\) and follow the condition \(l + m - j - k = nN \), where \(n\) is some integer. 

In the harmonic case where \(\Delta = -U/2 = 0\), it can be seen that in the momentum basis the detuning $\Delta - 2J\cos\left(\frac{2\pi k}{N}\right)$ is zero for modes with \(k = nN/4\) (where \(n\) is some integer). As such the drive is resonant to these modes, and these momenta determine the correlation profile. In the anharmonic case in turn  \(\Delta = -U/2\) and is large compared to \(2J\) and the mode with \(k = 0\) is closest to resonance with the drive.

\section{\label{sec:3}Conclusion}
In conclusion, we have considered nonlinear cavity arrays where the relaxation on the transition \(|m+1 \rangle \rightarrow |m \rangle\) is greater than the relaxation on the transition \(|n+1 \rangle \rightarrow |n \rangle\) for \(m > n\), and shown that these have a stationary state with similar properties as a Mott insulator if they are parametrically driven on the transition \(|0 \rangle \leftrightarrow |2 \rangle\) in each lattice site. We have also explored the transition to a delocalized phase with increasing tunneling rate and find that first order coherence does build up spontaneously. In contrast to the equilibrium case, this first order coherence does not become long range and decreases for very large tunneling rates after reaching a peak. We attribute this non-monotonic behavior to dephasing processes that become more important as the number fluctuations increase as a consequence of enhanced tunneling. In future research it would be interesting to corroborate these trends by exploring regimes with even larger tunneling rates $J$ (which difficult to access with our current numerics) and to investigate higher density Mott lobes for systems with a larger cascade of dissipation rates.

\bigskip 

\begin{acknowledgments}
The authors would like to thank Edmund Owen for discussions and helpful comments on the manuscript.
	O.T.B acknowledges support from EPSRC CM-CDT grant No.~EP/G03673X/1 and M.J.H. from EPSRC grant No.~EP/N009428/1.
\end{acknowledgments}

\end{document}